\documentclass[%
 aip,
 amsmath,amssymb,
 reprint,%
]{revtex4-1}

\usepackage{graphicx}
\usepackage{dcolumn}
\usepackage{bm}
\usepackage{upgreek}
\usepackage[utf8]{inputenc}
\usepackage[T1]{fontenc}
\usepackage{mathtools}
\usepackage{mathptmx,xcolor}
\usepackage{etoolbox}

\makeatletter
\def\@email#1#2{%
 \endgroup
 \patchcmd{\titleblock@produce}
  {\frontmatter@RRAPformat}
  {\frontmatter@RRAPformat{\produce@RRAP{*#1\href{mailto:#2}{#2}}}\frontmatter@RRAPformat}
  {}{}
}%
\makeatother
\begin{document}

\preprint{AIP/123-QED}

\title{3D tracking of particles in a dusty plasma by laser sheet tomography}
\author{Wentao Yu}%
 \email{wyu54@emory.edu.}
\author{Justin C. Burton}

\affiliation{ 
Department of Physics, Emory University, Atlanta, GA 30033, USA
}%

\date{\today}

\begin{abstract}
The collective behavior of levitated particles in a weakly-ionized plasma (dusty plasma) has raised significant scientific interest. This is due to the complex array of forces acting on the particles, and their potential to act as in-situ diagnostics of the plasma environment. Ideally, the three-dimensional (3D) motion of many particles should be tracked for long periods of time. Typically, stereoscopic imaging using multiple cameras combined with particle image velocimetry (PIV) is used to obtain a velocity field of many particles, yet this method is limited by its sample volume and short time scales. Here we demonstrate a different, high-speed tomographic imaging method capable of tracking individual particles. We use a scanning laser sheet coupled to a single high-speed camera. We are able to identify and track tens of individual particles over centimeter length scales for several minutes, corresponding to more than 10,000 frames. 
\end{abstract}

\maketitle

\section{\label{sec:level1}Introduction}
Dusty plasma, where particles are immersed in a weakly-ionized plasma, is ubiquitous in interstellar and planetary environments. It is also often encountered in industrial plasma processing. The immersed particles in a dusty plasma experience a wide range of environmental forces (i.e. electrostatic forces, drag from both neutral gas and ions), and environment-mediated interaction forces (forces from other particles), which can be non-pairwise and non-reciprocal in nature \cite{chaudhuri2011complex,lampe2000interactions,melzer2019physics,nambu2001effect,winske2001nonlinear}. As a result of this inherent complexity, systems of many particles exhibit a variety of emergent behaviors \cite{bockwoldt2014origin,Choudhary2020,gogia2017emergent,Gogia2020,Harper2020,ivlev2015statistical,Melzer2019b,qiao2015mode,Williams2007,himpel2012three}. Stereoscopic, multi-camera imaging is commonly used to obtain 3D information about particles' dynamics and kinetics \cite{melzer2016stereoscopic,mulsow2017analysis,thomas2004application}. A well-developed technology, particle image velocimetry (PIV), can be used to obtain a velocity field from multi-camera movies \cite{elsinga2006tomographic,raffel1998particle,williams2011application}. Calculating the velocity field can reveal spatial and temporal variations in the dynamics, but ultimately averages the individual particle dynamics over some multi-particle length scale. Yet many problems of scientific interest, for example, the mechanism of particle charging and interaction forces, are challenging to investigate without tracking individual particles over long times. 

Advanced particle tracking velocimetry (PTV) techniques that track individual particles have been mostly applied to the 2D motion of particles using a single camera \cite{feng2008solid,feng2016particle,zeng2022determination}. Some phenomena, for example, stochastic oscillations \cite{liu2009transverse} and spontaneous oscillations \cite{Harper2020} can be observed by analyzing individual particle motions from 2D PTV, but dynamical information perpendicular to the viewing plane is lost. 3D PTV by stereoscopic imaging is plagued by the challenge of delicate calibration of multiple cameras, identifying the same particle appearing in different cameras, and linking those particles over frames. Recent studies have approached this task by using statistical tools \cite{himpel2011three,himpel2014stereoscopy,schanz2016shake} or machine learning \cite{alpers20153d,himpel2021fast,wang2020microparticle}. Using statistical inference, the likelihood of a particle at a certain voxel is calculated from the brightness of its calibrated 2D projection in all the cameras, and also the likelihood of a particle existing at this pixel in the previous frame. With machine learning, light spots of particles are simulated by a pre-determined distribution and the model is trained on mapping the videos of multiple cameras to the simulated particle positions. Using these techniques, individual particles in dense dusty plasmas can be tracked for up to 30-50 frames in a volume of 10-100 cubic millimeters \cite{himpel2011three,himpel2012three,himpel2014stereoscopy,himpel2021fast,schanz2016shake}. 

In contrast to stereoscopic imaging, 3D tomography relies on a laser sheet that moves relative to the particles. If the position of the laser sheet is known in time, then a sequential series of images can be used to track the particles in three dimensions. In dusty plasmas, 3D tomography has mostly been used to examine static properties at scan rate of $\approx$ 1 Hz or less \cite{nefedov1998potential,karasev2009optical,zuzic2000three,thomas2003complex,dietz2018fcc}. Tomography methods has been extended to faster scanning rates, up to 15 Hz \cite{samsonov2008high,wang2020surface}, yet some experimental problems arise at these speeds. Most noticeably, the inertia of the oscillating mirrors used to deflect the laser limits the size and scanning speed of the imaging volume. Consequently, examining rapidly-moving dust particles for long periods of time remains a challenge. This is important for applications involving dynamical inference of the underlying forces driving the particles, or other dusty plasma phenomena outside of the well-studied Coulomb crystal state. 

Here we report a 3D tomographic imaging and tracking method that is conceptually simple and easy to implement. The method is suitable for tracking multiple particles in large volumes over long times. We use a scanning laser sheet with a single camera to obtain 3D trajectories of particles, saving the difficulties of calibration and identification with multiple mirrors or cameras. Since the particles in a dusty plasma are moving rapidly in time, we oscillate the height ($z$) of the illuminating 2D laser sheet with a shorter period than the characteristic time of the particle motion, up to a scanning rate of 500 Hz. As a result, particles at a certain $z$ only appear in specific frames of the camera. The images are then processed and particles are identified and tracked by Trackpy \cite{allan_daniel_b_2021_4682814} using a customized class. Subsequently, the trajectories are calibrated and adjusted for their sub-pixel accuracy. With this technique, we are able to track the 3D trajectories of 1-30 particles for 10,000 or more sequential frames in 1-10 cubic centimeters. We demonstrate this method on two distinct dusty plasma systems driven by vertical oscillations or magnetic field-induced ion flow. The mean particle positions, oscillation amplitudes, and characteristic frequencies of oscillation are reported for each individual particle.  

\section{Experimental design}
\begin{figure}[!]
\centering
\includegraphics[width=3.33 in]{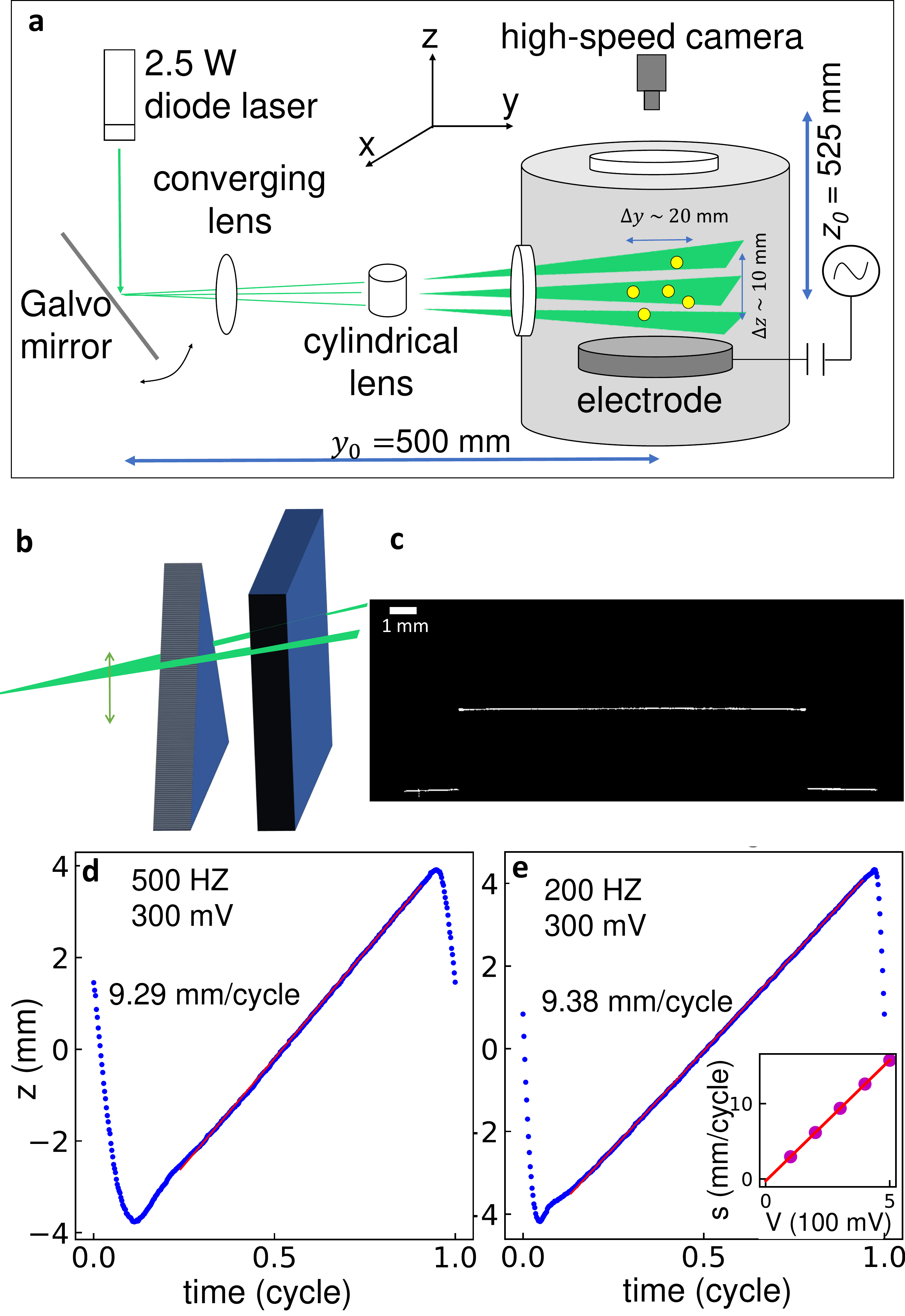}
\caption{(a) Experimental setup for the 3D tomographic imaging and particle tracking. The particles' scattered light from the oscillating laser sheet is imaged from above. (b) Sketch of a triangular screen placed in front of a rectangular screen used for calibrating the dynamics of the laser sheet height. (c) Example image showing the scattered light from the triangular screen (top white line). The two bottom lines are scattered from the rectangular screen. The calibrated height of the laser sheet with input sawtooth wave voltage = 300 mV and frequency = 500 Hz (d) and 200 Hz (e). Linear fits to the central region are shown with red lines. The slope $s$ of the linear fit versus input voltage with fixed frequency = 200 Hz is plotted in the inset of (e).}
\label{p1}
\end{figure}

Our experiments used melamine-formaldehyde (MF)
particles with diameters ranging from 8.0 to 12.8 $\upmu$m (microParticles GmbH). The particles were electrostatically levitated in a low-pressure, 13.56 MHz rf argon plasma above an aluminum electrode with a diameter = 150 mm (Fig.\ \ref{p1}a, similar to previous experiments \cite{gogia2017emergent, Harper2020,yu2022extracting,Gogia2020}). The rf plasma was driven with 0.3-5 W of power, and the gas pressure varied from 0.1-1 Pa. In this regime, the typical charge on a single particle ranged from 10,000-50,000 electrons. The particles were suspended at the edge of the plasma sheath approximately 1-2 cm above the aluminum electrode. The electrostatic confinement provided by the plasma sheath led to a near harmonic 3D trap. Upon displacement from its equilibrium position, a single particle experienced both vertical and horizontal oscillations, with a much higher frequency in $z$, as will be discussed in Sec.\ \ref{parallax}. Multiple particles in the same system experienced complex interactions often characterized by nonreciprocal forces \cite{ivlev2015statistical}.

To image and accurately track the particles' motion in 3D, a 2.5 W diode laser (Laserglow Technologies) was reflected by a mirror placed $y_0$ = 500 mm away from the center of the particles' motion and then expanded into a 2D sheet using a cylindrical lens (Fig.\ \ref{p1}a). The sheet was also focused in the $z$-direction by a plano-convex lens. The intensity profile of the laser sheet at the position of the particles was measured and well-fit by a Gaussian distribution with a standard deviation of 0.13 mm. The scattered light from the particles was captured by a high-speed v711 Phantom camera (Vision Research) located at $z_0 =$ 525 mm above the center of the particles. A galvo motor (Thor Labs), driven by a function generator (Agilent 33120A), was used to oscillate the mirror in a sawtooth wave pattern with a tunable amplitude and frequency. The input sawtooth wave frequency (100-500 Hz) determined the scanning frequency of the laser, and ultimately the 3D sampling rate of the particle motion.  We note that despite the high-intensity of the diode laser (2.5 W), when the laser is expanded into a sheet and oscillated vertically, the net force provided by the laser on individual particles was negligible.

To calibrate the height of the laser sheet at the center of the electrode during a single cycle of the input sawtooth wave, we built an isosceles right triangle out of stiff paper and placed it directly at the center of the imaging plane ($y=0$). We then placed a paper rectangle 4 mm behind the triangle (Fig.\ \ref{p1}b). Impinging light from the laser sheet was partially blocked by the triangle, leaving a bright line on the triangle whose length varied with height (Fig.\ \ref{p1}c, the upper line). The remaining unblocked laser sheet projected another line on the rectangle. The apparent length of the bright line on the rectangle varied by a small amount since it moved closer to and farther from the camera lens during a cycle (the lower line of Fig.\ \ref{p1}c). However, the actual length does not change with laser height, thus 
the bright line on the rectangle functioned as a fixed length scale at different $z$-positions so that we may calculate the length of the bright line on the triangle in millimeters. The corresponding height of the bright line on the triangle is then found using the geometry of the triangle since the angles are known. 

During the calibration, the camera frame rate was set to 19.95 times the input sawtooth wave frequency on the galvo motor, so 399 frames were captured over 20 cycles. The non-integer ratio of the frequencies led to a small drift so that different $z$-positions were sampled over many cycles, and those frames could be effectively shifted into one cycle. The calibrated $z$ position of the laser sheet in a cycle, as shown in Fig.\ \ref{p1}d-e, were near-sawtooth waves with a large area in the center of the cylce where $z$ increased linearly with time. The slope for the linear fit, $s$, did not sensitively depend on the frequency of the input sawtooth wave. However, the slope scaled linearly with the amplitude of the voltage driving the galvo, as shown in the inset of Fig.\ \ref{p1}e. With $y_0 = 500$ mm, $s = 3.2$ mm$\cdot$cycle$^{-1} \cdot$ per 100 mV.

In our experiments, we typically set the camera's recording frame rate to 20$\times$ the laser scanning frequency. This provides 20 vertical image slices per cycle.  The exposure time was always set at the maximum possible value with the given recording rate in order to capture as much of the scattered light as possible from the particles. We focused the camera sharply on the particles so that corresponding width of the scattered light spots on the image sensor is $\approx$ 2 pixels. We note that the depth of field of our camera lens was larger than the typical $z$ displacement of particles, so particles appeared in sharp focus despite their vertical motion. Each recorded movie was processed by Trackpy \cite{allan_daniel_b_2021_4682814}. A customized 3D-frame class, included in the supporting information, was used to group every 20 frames into a bundle, automatically detect from which frame a bundle began, and discard the first and last 2 frames in each bundle (they exist outside of the linear region in Fig.\ \ref{p1}d). The code, as provided in the supplemental information, produces a rudimentary database which contains the positions $x'$, $y'$, and $z'$ for each particle in each frame. After the initial tracking, the code also applies a SPIFF correction to alleviate small statistical errors in tracking due to pixel-locking \cite{yu2022extracting,yifat2017analysis}; a known and significant tracking issue that arises when the width of a particle light spot is comparable to the pixel size. 

\section{Laser divergence and parallax correction after tracking}

\label{parallax}

\begin{figure}[!]
\centering
\includegraphics[width=0.97\columnwidth]{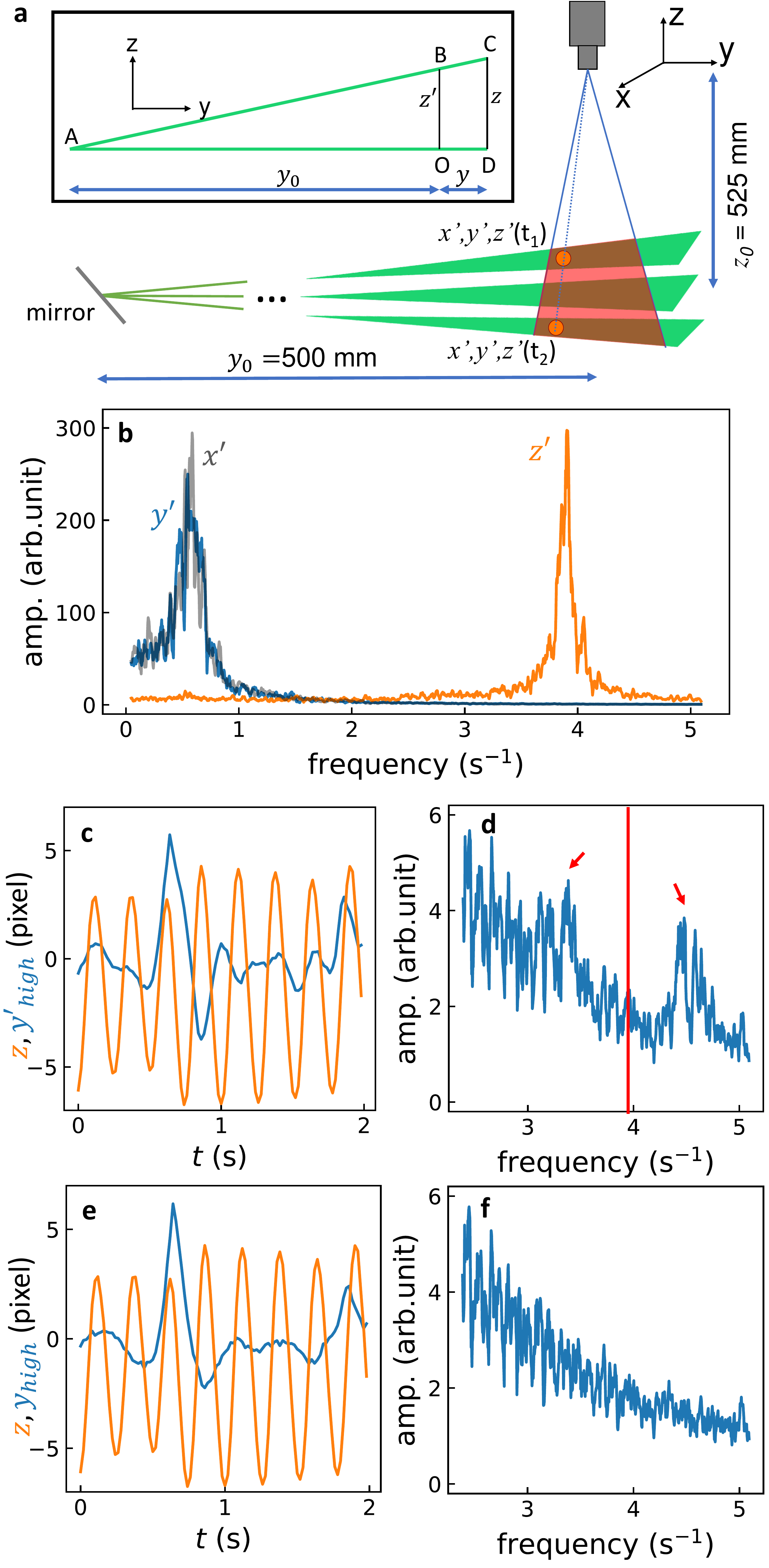}
\caption{(a) Diagram showing the intersection of the solid angle of view from the camera and the laser sheet (red shaded area). Two particles at different $x$ and $y$ may have the same $x'$ and $y'$ when viewed from the camera (the blue dotted line). Inset: the $y-z$ plane of view for the geometry of the imaging system. The mirror is at $A$. A particle located at $C$ will appear at the same height $z'$ as $B$. (b) The Fourier spectrum of $x'$ (gray), $y'$ (blue), and $z'$ (orange). The amplitude of $z'$ is magnified by 10x for clarity. (c) The high-frequency band, $y'_{high}$, of the $y'$ component from the rudimentary trajectory of a tracked particle shows a correlation with its $z$ position. (d) The Fourier spectrum of $y'$ has peaks at $\approx 3.3$ Hz and 4.5 Hz, indicated by the red arrows, which is related to the vertical oscillation frequency, 3.9 Hz, indicated by the red line. (e) After the micro-correction procedure, the high-frequency band, $y_{high}$, of the $y$ component from the same particle shows no correlation with $z$. (f) After the micro-correction, the Fourier spectrum of $y$ no longer displays a peak.}
\label{p2}
\end{figure}

In our experiments, gravity and electrostatic forces within the plasma sheath near the aluminum electrode are the largest forces exerted on the particles (both in the $z$-direction). These forces are approximately 10-100$\times$ larger than the horizontal confinement forces since the diameter of the electrode (15 cm) is much larger than the levitation height (1 cm). Additionally, the vertical confinement forces are much larger than typical particle interaction and drag forces, which is common for dust particles levitated in RF plasma sheaths \cite{gogia2017emergent,yu2022extracting}. As a result, particles experienced natural oscillations with frequencies $f_z = 4-25$ Hz in $z$ and $f_{xy} = 0.5-3$ Hz in the $xy$ plane. The wide range of frequencies come from two distinct experiments with different environmental conditions, as discussed in Sec.\ \ref{results}. Since particles move above and below the focal plane (Fig.\ \ref{p2}a), there is a small amount of imaging parallax, or coupling between the vertical and horizontal positions. The tracked, rudimentary trajectories with coordinates $x'$, $y'$, and $z'$ are different from the desired Cartesian coordinates $x$, $y$, and $z$. In one experiment, the natural frequency of oscillation in the $xy$ plane was $\approx$ 0.6 Hz, and the frequency in the $z$ direction was $\approx$ 3.9 Hz (Fig.\ \ref{p2}b). As a result of the imaging parallax, the high-frequency band ($>2.5$ Hz) in the Fourier transform of either $x'$ or $y'$ from single particles is highly correlated with oscillations in the $z$ position, as shown in Fig.\ \ref{p2}c. Two sideband peaks near 3.3 Hz and 4.5 Hz appear in the Fourier spectrum of $y'_{high}$, which stems from an amplitude modulation of the $y$ position due to $z$ oscillations, essentially mixing the signals (Fig.\ \ref{p2}d). 

To understand the origin of these sideband peaks, we first introduce a geometric, first-order linear correction used to handle the conversion between imaged, primed coordinates and the real Cartesian coordinates, denoted as a ``micro-correction'' in the tracking code. The micro-correction is applied with the following transformation:
\begin{eqnarray}
    z &=& z'\frac{y_0+y'}{y_0},\label{mc1}\\
    y &=& y'\frac{z_0-z}{z_0},\label{mc2}\\
    x &=& x'\frac{z_0-z}{z_0}.\label{mc3}
\end{eqnarray}
The first line corrects for the divergence of the laser sheet; particles imaged in the same frame may have slightly different actual $z$ positions (Fig.\ \ref{p2}a). As $y'/y_0\rightarrow 0$, this correction becomes negligible. Note that Eqs.\ \ref{mc1}-\ref{mc3} are derived purely from geometry. As shown in the inset of Fig.\ \ref{p2}a, point $A$ is the mirror, $O$ is the center of the particle system, and line $OB$ is the axis where the calibration of $z$ is conducted. A particle at position $C$ will appear in the same frame as a particle at point $B$. The measured $z'$ is $BO$, while its actual $z$ is $CD$. By trigonometry:
\begin{equation}
    \frac{z}{z'} = \frac{y+y_0}{y_0},
\end{equation}
which is equivalent to Eq.\ \ref{mc1}. Note that $y'$ is used instead of $y$ in the numerator of Eq.\ \ref{mc1}. However, this is a second-order effect $\mathcal{O}((y/y_0)^2)$, where $y/y_0\approx0.02$. Equations \ref{mc2} and \ref{mc3} represent a first order correction for imaging parallax, which is negligible as $z/z_0\rightarrow 0$. The derivation is identical to Eq.\ \ref{mc1}, except that the $y$ axis is replaced by $-z$ and $z$ is replaced by $y$.

To demonstrate why the sideband peaks are observed at 3.3 HZ and 4.5 HZ, respectively, we will assume a simplified model where a single particle's true trajectory is sinusoidal, given by:
\begin{eqnarray}
    y &=& A\cos(\omega_yt+\phi),\\
    z &=& B\cos(\omega_zt).
\end{eqnarray}
Due to parallax, as shown in the inset of Fig.\ \ref{p2}a, the measured $y$-position, $y'$ is 
\begin{eqnarray}
    y' &=& y + \frac{yz}{z_0},\\
    &=& y + \frac{AB}{z_0}\cos(\omega_yt+\phi)\cos(\omega_zt),\nonumber\\
    &=& y + \frac{AB}{2z_0}\left(\cos(\omega_zt-\omega_yt-\phi) + \cos(\omega_zt + \omega_yt+\phi)\right).\nonumber
\end{eqnarray}
Therefore, imaging parallax causes peaks in the Fourier spectrum of $y'$ at $\omega_z-\omega_y$ and $\omega_z+\omega_y$, which correspond to 3.3 HZ and 4.5 Hz.

 This linear correction is sufficient to remove the coupling between positions in the $xy$ plane and $z$, as shown in Fig.\ \ref{p2}e. The correlation between $z$ and $y_{high}$ (the high-frequency band of the corrected $y$ coordinate) has been removed. Also, the peaks at 3.3 Hz and 4.5 Hz no longer exist in the Fourier spectrum of $y$ after the correction (Fig.\ \ref{p2}f).  Finally, we note that since particles at different $z$ positions are recorded at slightly different times, the final time series of positions for a given particle may have non-uniform intervals. To obtain uniform time intervals between bundles, we use interpolation to the central time of each $z$ bundle. 

\section{Tracking results}

\label{results}

\begin{figure}[!]
\centering
\includegraphics[width=3.33 in]{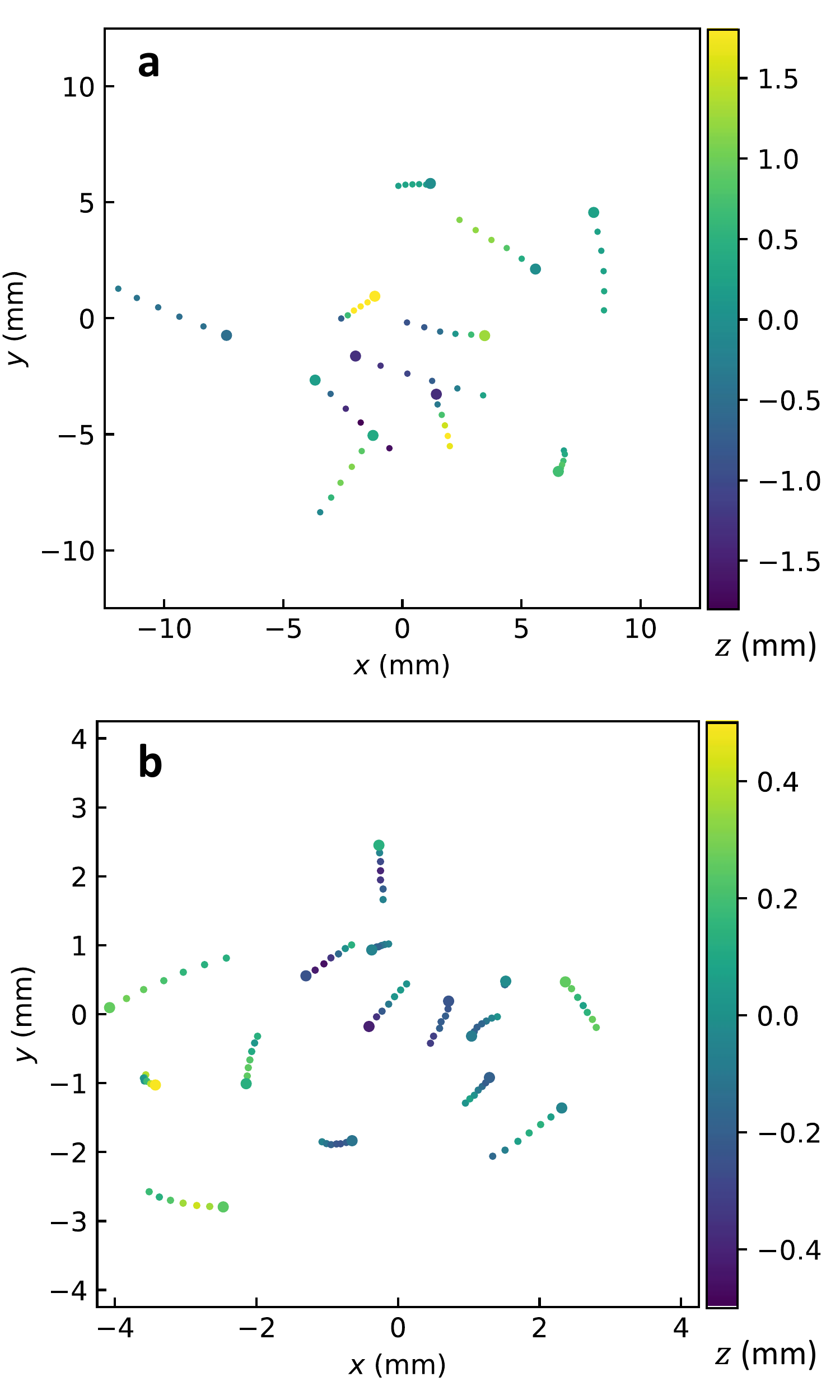}
\caption{(a) Snapshot of particle positions from a tracked, 11-particle system at 0.1 Pa. The $z$ position is indicated by the colorbar. The colored tails represent the positions in the previous 5 bundles. (b)  Snapshot of particle positions from a tracked, 15-particle system at 1.0 Pa. Here, the presence of the magnetic field leads to a much stronger confinement.}
\label{p3}
\end{figure}

\begin{figure}[!]
\centering
\includegraphics[width=\columnwidth]{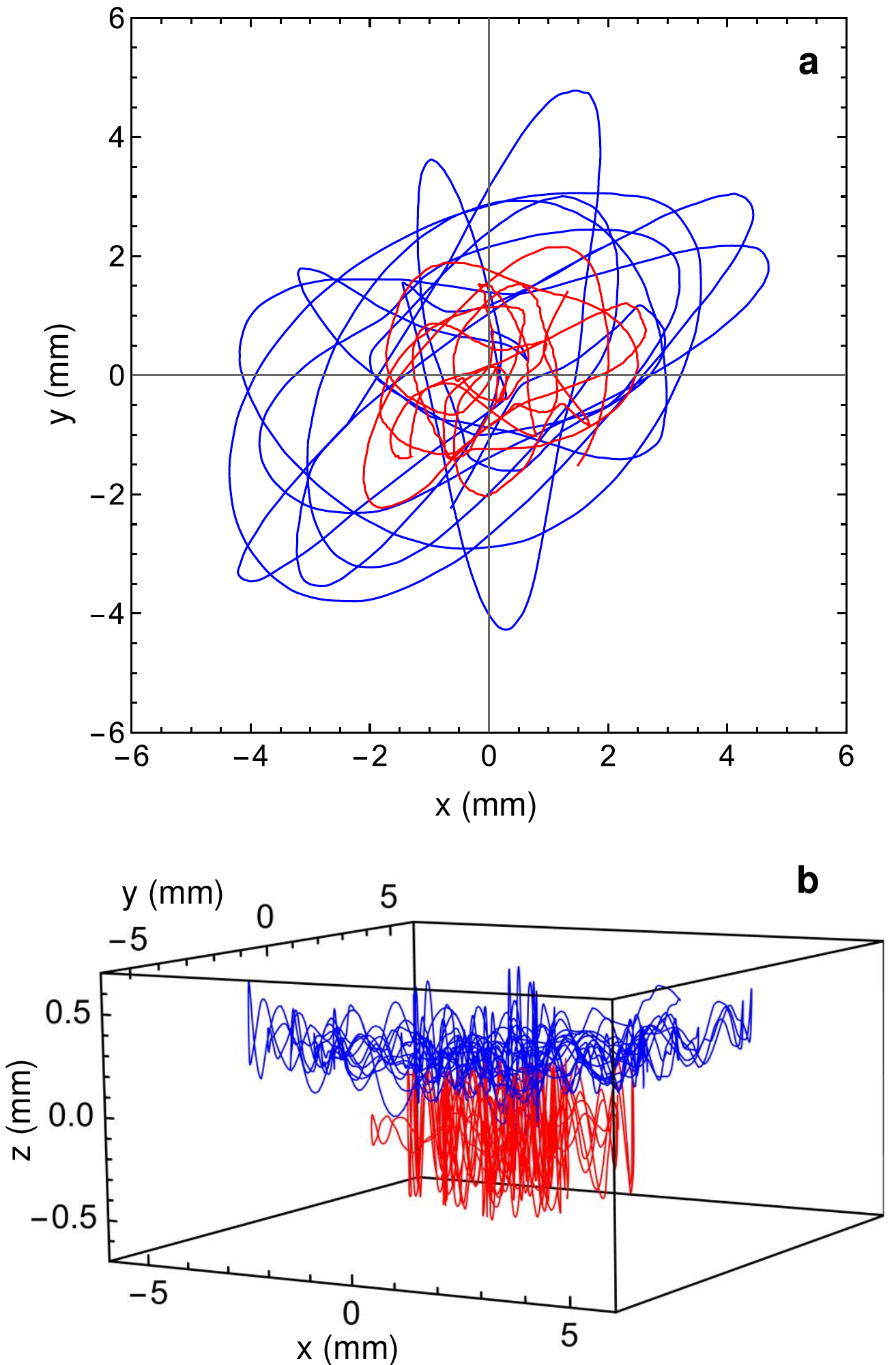}
\caption{(a) Projection in the $xy$ plane of the motion of 2 particles within a system of 15 particles. The system was driven by ion flow from a permanent magnet, as described in the text. The length of each trajectory is 5 s, and the sampling period is 0.005 s. (b)  The motion of the same 2 particles plotted in 3D. One particle (red) sits considerably lower in $z$ due to its larger mass. Both particles oscillate rapidly in the $z$-direction. In both panels, trajectories are connected by quadratic interpolation.}
\label{new_long}
\end{figure}

To demonstrate our imaging and tracking method, we used two collections of MF particles in distinct environmental conditions. At very low plasma pressure (0.1 Pa), we tracked an 11-particle system that was highly underdamped due to the low gas pressure. The particles experienced spontaneous vertical oscillations with amplitudes $\approx 1$ mm, which acted as a source of constant energy input into the system \cite{nunomura1999instability,Harper2020}. In the $xy$ plane, the particles moved with much larger amplitudes ($\approx 10$ mm). To track the system, the laser scanning frequency was set to 50 Hz and the camera frame rate was set to 1,000 Hz. The tracked particle positions at a single point in time are illustrated in Fig.\ \ref{p3}a. In a second experiment, we placed a large, 7.5 cm diameter rare-earth magnet within a cavity machined into the electrode. The vertical component of the magnetic field was measured to be $\approx$ 0.05 T at the levitation height of the particles. The gas pressure was set to 1.0 Pa. Under these conditions, the particles experienced rapid vortical motion in the $xy$ plane with a rotation period of $\approx$ 0.5 s. This voticity stems from an ion drag force since ions streaming towards the aluminum electrode have a finite velocity component in $x$ and $y$, and thus are deflected by the magnetic field \cite{samsonov1999instabilities,Choudhary2020}. The amplitudes of particle oscillations were $\approx 3$ mm in the $xy$ plane and $\approx 0.2$ mm in $z$. To track the system, the laser scanning frequency was set to 200 Hz and the camera frame rate was set to 4,000 Hz. The tracked particle positions at a single point in time are illustrated in Fig.\ \ref{p3}b. For both systems, we successfully and continuously tracked more than 10,000 frames without confusing particle positions or losing particles. Sample movies from these systems can be found in the supplementary information (Movie S1 and Movie S2). 

With the advantage of high-speed, 3D tracking, we can follow the motion of individual particles in a many body system over long times. In Fig.\ \ref{new_long} we show the trajectories for 2 particles over 5 s (1000 frames) from the second system, where particles experience vortical motion due to the external magnetic field. Figure \ref{new_long}a depicts just the motion within the $xy$ plane, which could be captured with traditional 2D particle tracking methods. One particle experiences a much larger range of motion (blue), whereas the other particle is more tightly confined to the center (red). In 3D, we see that this is because of their difference in mass, and thus their corresponding vertical positions (Fig.\ \ref{new_long}b). The particles are separated by $\approx 0.5$ mm in $z$. Both particles experience rapid vertical oscillations that are not observable in the 2D projection.

\begin{figure}[!]
\centering
\includegraphics[width=\columnwidth]{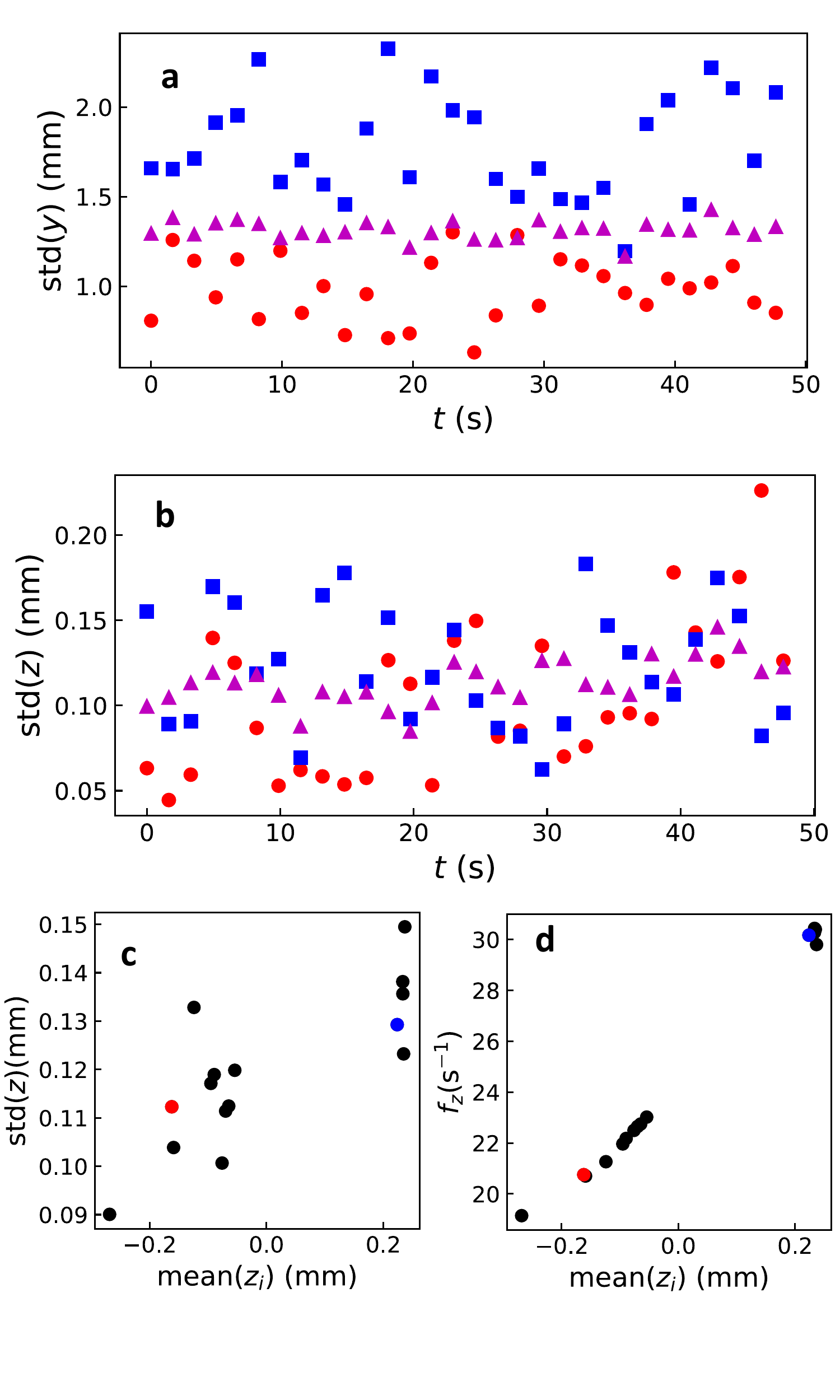}
\caption{The oscillation amplitude in $y$ (a) and $z$ (b) for two particles (red circles, blue squares); the same as shown in Fig.\ 4. The amplitude is computed using the standard deviation of positions in binned time windows of width 1.65 s (330 frames), which is 3x the horizontal period. The magenta triangles are the average amplitude for all 15 particles.  The oscillation amplitude (c) and peak frequency (d) in $z$ are calculated for each of the 15 particles over the whole time series. The two particles shown in Fig.\ 4 are colored blue and red, respectively.}
\label{p5}
\end{figure}

Despite attempts to use identical particles when conducting experiments, the manufacturing process always leads to a degree of heterogeneity. By tracking particles in 3D, we can harness the particles' heterogeneity to investigate their local plasma environment. Figure \ref{p5}a shows that particles that resided higher in the sheath (blue) always moved horizontally with a larger amplitude than average (magenta), and the particles that resided lower in the sheath (red) always moved with smaller amplitude. Here the amplitude is defined as the standard deviation (std) of the particle position over some time window. However, the amplitude of vertical oscillations was more intermittent with longer timescales, but generally showed the the same behavior (Fig.\ \ref{p5}b). For example, although the bottom (red) particle moved less in $z$ than average, during the last 5 s of the experiment, it experienced a larger amplitude of oscillation in $z$ (trajectories for these particles are shown in Fig.\ \ref{new_long}b). 

When averaged over the entire time series, the standard deviation of motion in $z$ increased with the mean $z$ position, meaning that lighter particles residing higher in the sheath experienced larger oscillations (Fig.\ \ref{p5}c). However, the vertical oscillation frequency ($f_z$) of each particle was most strongly correlated with the mean $z$ position, as shown in Fig.\ \ref{p5}d. Smaller particles sat higher in the sheath, and had a higher oscillation frequency. In principle, one can obtain the variation in the vertical electric field from the variation in $f_z$. Yet in practice, both the electric field and particle charge vary with $z$, which complicates this inference \cite{Harper2020}. By considering particle interactions, one may use the 3D trajectories to infer pairwise forces between particles (including estimates of the charge and mass) or predict their future motion, which may be further used to infer forces from the plasma environment. Finding low-dimensional representations of dynamics or new physical laws from many-body trajectories using machine learning is a rapidly expanding field of research \cite{champion2019data,gnesotto2020learning,rudy2017data,daniels2015automated,brunton2016discovering}. Our tracking method, combined with a machine learning-based algorithm, could potentially lead to a better understanding of the complex dynamics of dusty plasmas. These are ongoing efforts that we are pursuing.

\section{Conclusions}

3D tomographic imaging and tracking of dusty plasmas provides new ways of studying dust dynamics over large, ergodic time scales. In particular, detailed information about the position and acceleration of each particle can reveal information about the spatial dependence of environmental and interaction forces between particles.
The method is complementary to more well-developed, stereoscopic imaging methods, where smaller imaging volumes are examined over very short periods of times ($\approx 100$ frames). Stereoscopic methods are appropriate for much denser dust systems, and do not require a very high-speed camera capable of thousands of frames per second. Although we have only focused on systems with 10-20 particles here, our tomography method can be scaled up to many more particles, however, not without some challenges. The main restrictions to tracking larger numbers of particles involve temporal and spatial resolution. The camera and corresponding laser sheet frequency must be fast enough to capture oscillations in $z$ and close encounters between particles, where forces and accelerations can be large. Additionally, during a close encounter, where two or more particles come within 2 voxels, Trackpy may confuse the identity of those particles. Trackpy first identifies particles in each frame separately before linking all the located records. Thus during particle identification, information from the previous and next frame is not used. This problem could be improved with more advanced tracking algorithms, for example, the shake-the-box algorithm \cite{schanz2016shake}. Furthermore, statistical information such as the average brightness, average kinetic energy, and average $z$ position can potentially be used to better identify individual particles and link them before and after a close encounter.

\section*{Supplementary Material}
The supplementary material consists of a text file containing the code used to track the particles, in conjunction with TrackPy \cite{allan_daniel_b_2021_4682814}. There are also two movies (S1 and S2) that show the results of the particle tracking. The movies display the three-dimensional positions of particles in two separate experiments.

\begin{acknowledgments}
We wish to acknowledge fruitful discussions with Guram Gogia, Eslam Abdelaleem, and Ilya Nemenman. This material is based upon work supported by the National Science Foundation under Grant No. 2010524, and the U.S. Department of Energy, Office of Science, Office of Fusion Energy Sciences program under Award Number DESC0021290. The data that support the findings of this study are available from the corresponding author upon reasonable request.
\end{acknowledgments}

\nocite{*}
\bibliography{3D_tracking_paper}

\end{document}